# Short term period variable stars observed at OAUNI

# Estrellas variables de corto periodo observadas en el OAUNI

## Antonio Pereyra[1,2]* y Julio Tello[2,3]

[1] *Instituto Geofísico del Perú, Área Astronomía, Badajoz 169, Ate, Lima, Perú*

[2] *Facultad de Ciencias Universidad Nacional de Ingeniería (UNI), Av. Túpac Amaru 210, Rímac, Lima, Perú*

[3] *Grupo Astronomía, UNI, Av. Túpac Amaru 210, Rímac, Lima, Perú*

**RESUMEN**

Presentamos los primeros resultados científicos del programa de estrellas variables de corto periodo observadas en el Observatorio de la Universidad Nacional de Ingeniería (OAUNI) en los Andes peruanos. Estos resultados incluyen curvas de luz de buena calidad de estrellas delta Scuti, estrellas rápidamente oscilantes, así como binarias eclipsantes y cataclísmicas. La precisión fotométrica alcanzada por el instrumental disponible, y utilizada en los distintos subprogramas científicos, ha cumplido con las expectativas iniciales.

*Palabras Clave:* estrellas variables, delta Scuti, rápidamente oscilantes Ap, binarias cataclísmicas, binarias eclipsantes

**ABSTRACT**

We present the first scientific results of the program on short term period variable stars observed using the OAUNI facility at the peruvian Andes. These results include good quality light curves of delta Scuti stars, rapidly oscillating stars along with eclipsing and cataclysmic binaries. The photometric precision reached by the available instrumental and equipment, and used in the several scientific subprograms, has satisfied the initial expectations.

---

**\*** Corresponding author.:
E-mail: apereyra@igp.gob.pe

# 1 INTRODUCTION

The astronomical observatory of The National University of Engineering (OAUNI [1,2]) operates at the peruvian central Andes (Huancayo, 3300 m.u.s.l.) since 2015. This ongoing effort aims to provide of a facility to develop science programs, teaching and outreach in astronomy. Between its prioritized scientific programs is the monitoring of short term period variables stars. This program was chosen to test the OAUNI capabilities to gather complete light curves for objects with periods less than one night (up to eight hours) and to test stellar microvariability (periods in order of minutes) in selected astrophysical objects.

The OAUNI program of short term period variable stars includes several types of systems. Special attention was given to pulsating variables and binaries systems with periods of hours or less. Relatively bright systems with brightness variation in order of tenths of magnitudes were initially tested with success. Systems with variation in hundredth of magnitudes were more challenging to the combination of the OAUNI telescope and detector [3].

In the following of this section, a short review of types the short term period variables gathered with OAUNI will be presented in order to put the astrophysical objects in context.

## 1.1 Pulsating stars

Pulsating variables are stars that show periodic variations of their surface layers in expansion and contraction [4]. The pulsation is radial when the star remains spherical in shape, and non-radial when exists a periodic deviation from a sphere. The types of pulsating variables may be classified by the pulsation period and the shapes of their light curves. The last ones are consequence of the mass and evolutionary status of the star. When the period is of order of few hours up to tens of days, the star is called as a short period pulsator. This group includes the Cepheids, RR Lyrae stars, and delta Scuti stars and occupies the instability strip in the Hertzsprung-Russell (HR) diagram [5].

### 1.1.1. delta Scuti stars

The delta Scuti stars are young pulsating stars of approximately two solar masses. In the HR diagram, they are situated where the instability strip crosses the main sequence and central hydrogen or shell hydrogen burning happens. The pulsations are a consequence of the supplied energy in the internal ionizations zones of elements as helium. Both radial and non-radial oscillations occur in delta Scuti stars. Those are generally low order pressure modes with periods between tens of minutes and several hours. The observed amplitudes span from thousandths up to tenths of a magnitude [6].

### 1.1.2. Rapidly oscillating Ap stars

The rapidly oscillating Ap (roAp) stars are also situated on the instability strip, close to the delta Scuti stars locus in the HR diagram. The roAp stars are A-type stars with strong magnetic fields and peculiar chemical surface composition. About forty roAp stars are known with photometric variability up to hundredths of magnitude and periods below than twenty minutes. The oblique pulsator model is used to explain the observed multiplets in the roAp pulsation modes, where it is necessary a misaligned magnetic axis with respect to the rotation axis [7, 8].

## 1.2 Cataclysmic variables

Cataclysmic variables (CV) are interacting binary stars with transferring mass of a normal star to its companion. This process yields occasional energetic outburst events with a strong mutual influence between their components. The canonical model is represented by a white dwarf star that is accreting material from a lower mass and very close red star with an accretion disk formed around the first one [9].

### 1.2.1 Intermediate polar

An Intermediate Polar is a CV with a white dwarf and a cool main-sequence secondary star. In these systems the canonical model requires that the inner disk is disrupted by the magnetic field of the rapidly rotating white dwarf with rotational periods range from tens of seconds to few hours [10]. Approximately 1% of the known CVs are confirmed as intermediate polar stars [11].

## 1.3 Eclipsing binaries

Eclipsing binaries are systems of stars with an orbital plane lying near the line-of-sight of the observer. This configuration lets the components eclipse one another with an apparent and periodic brightness decrease. The period of the eclipse spans from minutes to years.

### 1.3.1 W Ursae Majoris variable

A W Ursae Majoris variable is a type of eclipsing binary variable star also known as a low mass contact binary. In these systems both components share a common envelope between the inner and outer contact surfaces, and are therefore in physical contact with one another, leading to luminosity and mass exchange. Such systems show light curves with very similar eclipse depths and little color variation over the orbital cycle [12].

## 2 OBSERVATIONS AND REDUCTIONS

The observations were gathered using the 0.5m OAUNI telescope [2] installed at the Huancayo site of the Geophysical Institute of Peru (IGP in spanish). The observations reported here were performed between 2016 and 2018 mainly during the dry season at the central peruvian Andes. The used detector was a front-illuminated CCD STXL-6303E (manufactured by SBIG[1]) of 3072×2048 pixels2 and 9μm/pixel. This detector along with the focal ratio f/8.2 of the optical system yields a plate scale of 0.45"/pixel and a field-of-view (FOV) of 23'×15'.

In this work we present the preliminary reduction of one star as example in each subprogram of the OAUNI short period term variability monitoring. All the images were reduced using the IRAF[2] environment with the typical corrections of dark current and flat field. Special own IRAF scripts were used to align the images and DAOPHOT aperture photometry was extensively used.

The results presented in this work confirm previous photometric studies about the achieved precision with the OAUNI detector [3]. We used the differential photometry technique to construct all the light curves showed in this work, where the instrumental magnitude of the interest object is discounted by the instrumental magnitude of one comparison star in the same field of view. Another comparison star is always measured as double checked. The large FOV of our detector facilitates to choose the better comparison stars in each case.

The log of observations is shown in Tables 1 and 2 for the years 2016–2017 and 2018, respectively. The observation date is indicated in Col. (1) using the format *yyyymmdd* with the observed object in Col. (2). The type of short term period variable star in shown in Col. (3) with five types: intermediate polar, cataclysmic, delta Scuti, eclipsing and rapidly oscillating Ap star. The filter used in each observation is indicated in Col. (4) with the number of measurements in a given observation sequence in Col. (5) and the integration time by individual measurement in Col. (6). Fig. (1) Indicates the observed hours during the 2016–2018 campaigns distributed in each subprogram of our short term period variables monitoring.

Between the pulsating variables our program has monitored up now six delta Scuti stars (BP Peg, CY Aqr, DY Peg, EH Lib, V567 Oph, and YZ Boo) and seven roAp stars (HD 122970, HD 143487, HD 185256, HD 196470, HD 213637, HD 217522, and LZ Hya). The total of individual measurements for our delta Scuti sample is approximately of 3900 observations and for the roAp sample more than 6200 observations between the campaigns of 2017 and 2018. Figs. 2 and 3 shows the distribution of observed hours for the delta Scuti and roAp stars observed up now.

Considering the sample of cataclysmic variables an extensive monitoring was performed on the intermediate polar variable FO Aqr with eleven observed nights and more than 2700 individual measurements spanned in three years (2016–2018). Other cataclysmic variables observed on 2017 campaign were five dwarf novae (V893 Sco, CTCV J2118-3412, EI Psc, GW Lib, and V0701 Tau) adding more than 1260 measurements.

With respect to the eclipsing binaries, our program monitored a couple of objects, one W Uma-type (V357 Her) and HW Vir-type (NSVS 4256825) with more than 1200 measurements.

## 3 ANALYSIS

### 3.1 delta Scuti stars

Our delta Scuti monitoring program have observed more than 25 hours (Fig. 1) distributed in six objects (Fig. 2). In the following, we analyzed one of them (CY Aqr) in order to determine its period directly of our own data.

#### 3.1.1 CY Aqr

CY Aqr is a short-period (P = 87.9 min) delta Scuti star with a large-amplitude variability (between 10.4 and 11.1 mag in V). For more the eighty years since its discovery [13], this pulsating star has been extensively observed, and several investigations of its changing pulsation have been published [14,15,16,17,18,19].

Fig. 4 shows the FOV gathered with OAUNI around CY Aqr on 2017/06/22 stacking the 3.3 hours sequence observed on this night. These measurements were obtained using a R broadband filter. The CY Aqr light curve is shown in Fig. 5 displaying approximately the 2.25×period observed. We applied the Lomb-Scargle periodogram [20] to the CY Aqr light curve using the NASA Exoplanet Archive online application[3]. Fig. 6 shows that the main power is centered in P = 0.061108444 days or 87.996 minutes in striking accord with the Cy Aqr period found in the literature. The CY

---

[1] *Santa Barbara Group - Diffraction Limited, diffractionlimited.com*
[2] *IRAF is ditributed by the National Optical Astronomy Observatory, which is operated by the Association of Univerities for Research in Astronomy, Inc., under cooperative agrément with the National Science Foundation*

[3] *https://exoplanetarchive.ipac.caltech.edu/cgi-bin/Pgram/nph-pgram*

Aqr diagram phase using this period is also shown in Fig. 6.

TABLA 1. Observation log 2016-2017

| Date | Object | type | filter | N | IT (s) |
|---|---|---|---|---|---|
| 20160714 | FO Aqr | int. polar | R | 30 | 20 |
| 20160714 | FO Aqr | int. polar | R | 150 | 20 |
| 20160802 | FO Aqr | int. polar | R | 300 | 20 |
| 20160803 | FO Aqr | int. polar | R | 320 | 20 |
| 20160804 | FO Aqr | int. polar | R | 300 | 20 |
| 20160805 | FO Aqr | int. polar | R | 91 | 20 |
| 20160811 | FO Aqr | int. polar | R | 25 | 20 |
| 20170525 | V893 Sco | cataclysmic | V | 243 | 20 |
| 20170622 | CY Aqr | delta Scuti | V | 471 | 20 |
| 20170623 | V567 Oph | delta Scuti | V | 141 | 20 |
| 20170624 | V567 Oph | delta Scuti | V | 387 | 20 |
| 20170628 | V357 Her | delta Scuti | R | 400 | 20 |
| 20170630 | V893 Sco | cataclysmic | V | 279 | 20 |
| 20170722 | EH Lib | delta Scuti | R | 45 | 20 |
| 20170722 | EH Lib | delta Scuti | R | 284 | 20 |
| 20170723 | EH Lib | delta Scuti | R | 330 | 20 |
| 20170723 | EH Lib | delta Scuti | R | 45 | 20 |
| 20170726 | YZ Boo | delta Scuti | R | 331 | 20 |
| 20170727 | YZ Boo | delta Scuti | R | 350 | 20 |
| 20170727 | YZ Boo | delta Scuti | R | 50 | 20 |
| 20170728 | YZ Boo | delta Scuti | R | 350 | 20 |
| 20170728 | DY Peg | delta Scuti | R | 268 | 20 |
| 20170729 | CTCV J2118-3412 | cataclysmic | V | 334 | 20 |
| 20170730 | V357 Her | delta Scuti | R | 336 | 20 |
| 20170730 | HD 217522 | roAp | R | 58 | 6 |
| 20170730 | HD 217522 | roAp | R | 123 | 4 |
| 20170730 | HD 217522 | roAp | R | 40 | 3 |
| 20170730 | HD 217522 | roAp | R | 33 | 2 |
| 20170730 | HD 217522 | roAp | V | 450 | 2 |
| 20170730 | EI Psc | cataclysmic | R | 128 | 20 |
| 20170731 | BP Peg | delta Scuti | R | 450 | 20 |
| 20170819 | NSVS 14256825 | eclipsing | V | 450 | 20 |
| 20170819 | NSVS 14256825 | eclipsing | V | 82 | 20 |
| 20170820 | HD 213637 | roAp | V | 133 | 4 |
| 20170820 | HD 213637 | roAp | V | 241 | 10 |
| 20170821 | HD 213637 | roAp | V | 160 | 15 |
| 20170821 | HD 213637 | roAp | V | 97 | 15 |
| 20170822 | HD 213637 | roAp | V | 183 | 12 |
| 20170823 | GW Lib | cataclysmic | V | 1 | 20 |
| 20170823 | V893 Sco | cataclysmic | V | 6 | 20 |
| 20170823 | V0701 Tau | cataclysmic | V | 270 | 20 |
| 20170918 | DY Peg | delta Scuti | R | 154 | 20 |
| 20170918 | DY Peg | delta Scuti | R | 250 | 20 |

TABLA 2. Observation log 2018

| Date | Object | type | filter | N | IT (s) |
|---|---|---|---|---|---|
| 20180516 | LZ Hya | roAp | V | 199 | 20 |
| 20180517 | LZ Hya | roAp | V | 100 | 20 |
| 20180517 | LZ Hya | roAp | V | 200 | 20 |
| 20180615 | HD 143487 | roAp | V | 36 | 20 |
| 20180615 | LZ Hya | roAp | V | 14 | 20 |
| 20180615 | HD 143487 | roAp | V | 173 | 20 |
| 20180706 | LZ Hya | roAp | V | 300 | 20 |
| 20180706 | FO Aqr | int. polar | R | 300 | 20 |
| 20180706 | FO Aqr | int. polar | R | 51 | 20 |
| 20180707 | LZ Hya | roAp | V | 261 | 20 |
| 20180707 | FO Aqr | int. polar | R | 179 | 20 |
| 20180707 | FO Aqr | int. polar | R | 125 | 20 |
| 20180708 | HD 122970 | roAp | V | 300 | 8 |
| 20180708 | HD 122970 | roAp | V | 168 | 8 |
| 20180708 | FO Aqr | int. polar | R | 300 | 20 |
| 20180708 | FO Aqr | int. polar | R | 136 | 20 |
| 20180708 | HD 196470 | roAp | V | 80 | 20 |
| 20180709 | HD 185256 | roAp | V | 86 | 20 |
| 20180709 | HD 185256 | roAp | V | 31 | 20 |
| 20180709 | HD 185256 | roAp | V | 147 | 20 |
| 20180713 | HD 185256 | roAp | V | 198 | 20 |
| 20180713 | FO Aqr | int. polar | R | 300 | 20 |
| 20180713 | HD 185256 | roAp | V | 200 | 20 |
| 20180715 | HD 143487 | roAp | V | 50 | 20 |
| 20180715 | HD 143487 | roAp | V | 80 | 15 |
| 20180715 | HD 122970 | roAp | V | 200 | 8 |
| 20180715 | HD 122970 | roAp | V | 101 | 8 |
| 20180715 | HD 143487 | roAp | V | 300 | 15 |
| 20180715 | HD 196470 | roAp | V | 300 | 15 |
| 20180716 | HD 143487 | roAp | V | 161 | 15 |
| 20180716 | HD 122970 | roAp | V | 300 | 6 |
| 20180716 | HD 143487 | roAp | V | 220 | 10 |
| 20180716 | HD 143487 | roAp | V | 250 | 15 |
| 20180716 | FO Aqr | int. polar | R | 171 | 20 |
| 20180716 | HD 196470 | roAp | V | 300 | 15 |

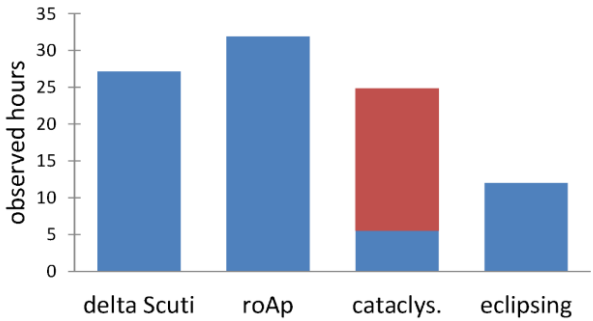

Figure 1. Distribution of observed hours in 2016–2018 OAUNI campaigns for the subprograms associated to the monitoring of short term period variables. In red are the observed hours for the FO Aqr monitoring.

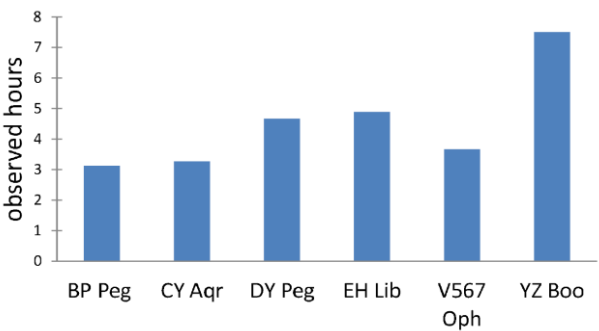

Figure 2. Distribution of hours in delta Scuti stars observed at OAUNI in 2016–2018 campaigns.

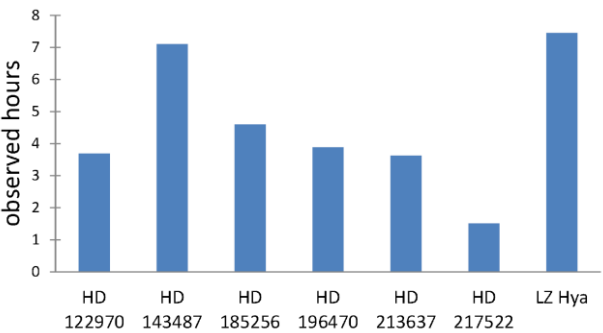

Figure 3. Distribution of hours in roAp stars observed at OAUNI in 2016–2018 campaigns.

### 3.2 roAp stars

The OAUNI roAp monitoring program have observed more than 31 hours (Fig. 1) distributed in seven objects (Fig. 3) during the 2017–2018 campaigns. In the following, we analyzed one of them (HD 217522) in order to determine the presence of a hidden period.

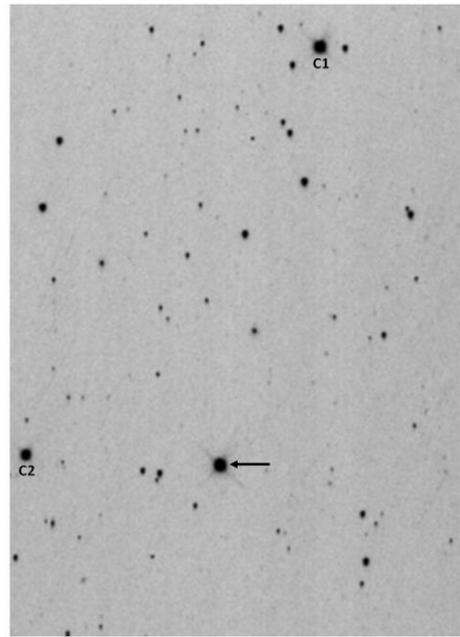

Figure 4. OAUNI stacked image (471 images×20s = 3.3h, including overhead) of CY Aqr observed on 2017/06/22. North in top and East is left. The FOV shown is 11.4'×16.1'. The position of CY Aqr is highlighted (arrow) along with the comparison stars (C1 and C2).

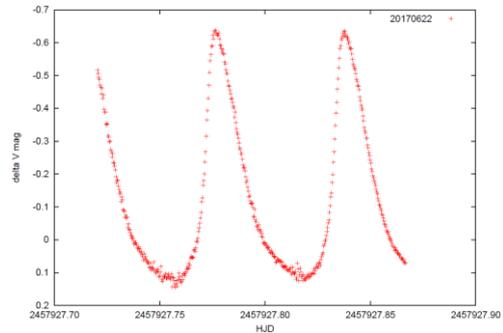

Figure 5. Light curve for CY Aqr on 2017/06/22.

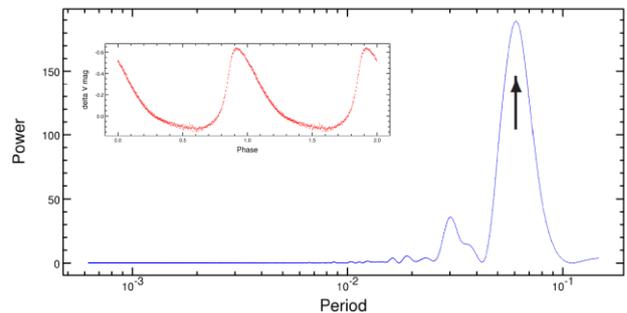

Figure 6. CY Aqr periodogram showing the main found period P = 0.061108444 days = 87.996 minutes. The insight shows the diagram phase using the found period.

#### 3.2.1 HD 217522

HD 217522 is a roAp star which earlier photometric observations revealed only one pulsation frequency$_1$ = 1.21509 mHz (or period = 13.7 min, [21]). Additional observations found a well-defined second frequency$_2$ = 2.0174 mHz (or period = 8.3 min, [22,23]). Interestingly, HD 217522 shows amplitude modulation over a time-scale of the order of a day, much shorter than what has been observed in other roAp stars [23].

The OAUNI roAp program observed HD 217522 on 2017/07/30 using the V broadband filter. Figure 7 shows the field of view around HD 217522 gathered on that night. The total observed sequence was 450 images of two seconds each one. Considering the overhead, this object was monitored by 1.1 hour, enough time to obtain any of the periods found in the literature.

Figure 8 shows the HD 217522 light curve observed on 2017/07/30 where the rapid oscillation in magnitude is very evident. Interestingly the mean photometric error in our differential photometry (0.004 mag) by individual measurement is at least one order of magnitude less than the typical dispersion peak-to-peak observed (0.029 mag). This condition is important to have chance in finding hidden periods associated to variations in amplitude of mmag using periodograms.

Figure 9 shows the Lomb-Scargle periodogram for the HD 217522 light curve. The first two higher peaks do not represent real periods associated to our object and probably are artifacts produced by short size of our interval time of measurements. The third peak is P = 13.8 min in good accord with the first fundamental period found for HD 217522 in the literature [21]. The fourth peak in Fig. 10 is P = 10.6 min and does not seem associated to the second fundamental period [22].

### 3.3 Cataclysmic binaries

Our cataclysmic variables monitoring program have observed approximately 25 hours (Fig. 1), being that 20 hours were used on the FO Aqr star. In the following, we analyzed four of the eleven observed nights (Tables 1 and 2) available for this object. This will be used to determine the periods claimed in the literature for FO Aqr.

#### 3.3.1 FO Aqr

FO Aqr is an intermediary polar with a system orbiting period of 4.85 hours [24]. The system underwent a low state in 2016 (decrease of the optical luminosity by 2 mag, from a normal state with V = 13.5 to a faint state with V = 15.5), from which it recovered slowly and steadily over a time scale of several months [25]. The system has a very strong optical pulsation of 20.9 minutes, associated with the rotational period of the accreting white dwarf [26]. In the 2016 lowest state, a very prominent 11.26 minute was discovered [27], corresponding with one-half of the beat period between the spin and orbital periods. After this episode the system come back slowly to its normal high state brightness, indicating a temporary drop-off in the mass-transfer rate between the two stars [28].

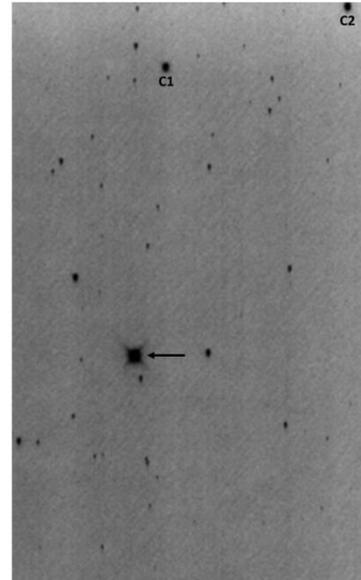

Figure 7. OAUNI stacked image (450 images×2s = 1.1 hour, including overhead) of HD 217522 observed on 2017/07/30. North in top and East is left. The FOV shown is 13.0'×21.5'. The position of HD 217522 is highlighted (arrow) along with the comparison stars (C1 and C2).

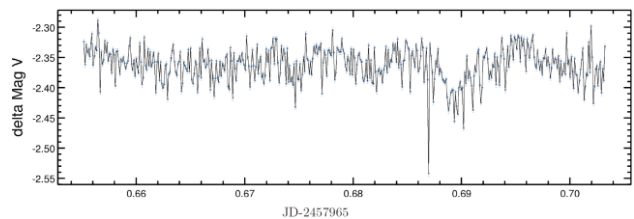

Figure 8. Light curve of HD 217522 on 2017/07/30.

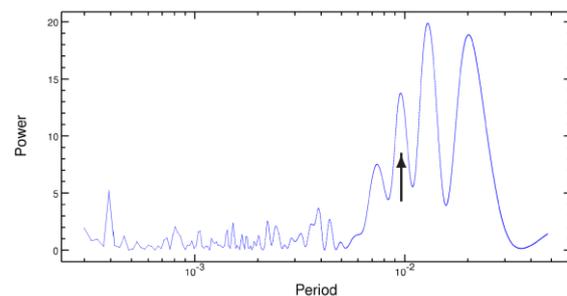

Figure 9. Periodogram of HD 217522 for 2017/07/30 data. The peak indicated is the period P = 13.8 min in accord with the literature.

Our first observations of FO Aqr were during the 2016 low state. New data were collected in 2018 campaign (July) presumably in an another low state [29]. Figure 10 shows the FOV gathered with OAUNI around FO Aqr and it corresponds to an 2.1 hours stacked sequence observed on 2016/08/02. A R broadband filter was used in this monitoring. Figure 11 (top) shows the FO Aqr light curve in four observed nights in 2018/July. The bottom image highlights the fast variability of the system in each night. In particular, the light curves for the second and third night present the longest time sequence letting a better analysis of the periods. Finally, Figure 12 shows the Lomb-Scargle periodogram for all the four light curves indicating the two periods observed previously in the literature. The main period with the higher power ($P_1$ = 20.58 minutes) corresponds as the one observed usually in the high state. The second peak ($P_2$ = 11.25 minutes) corresponds as the new period observed in the 2016 low state [27].

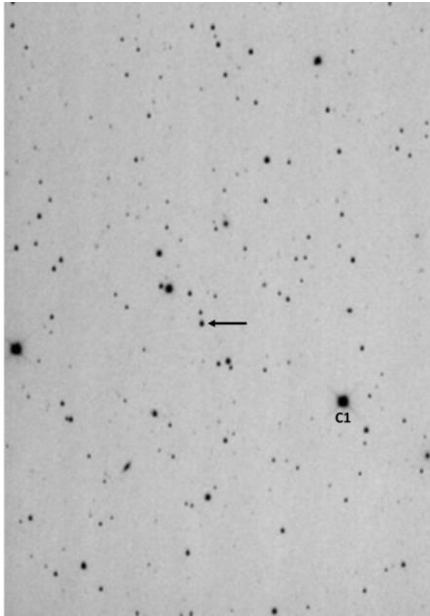

Figure 10. OAUNI stacked image (300 images×20s = 2.1h, including overhead) of FO Aqr observed on 2016/08/02. North in top and East is left. The FOV shown is 13.9'×19.0'. The position of FO Aqr is highlighted (arrow) along with the comparison star (C1).

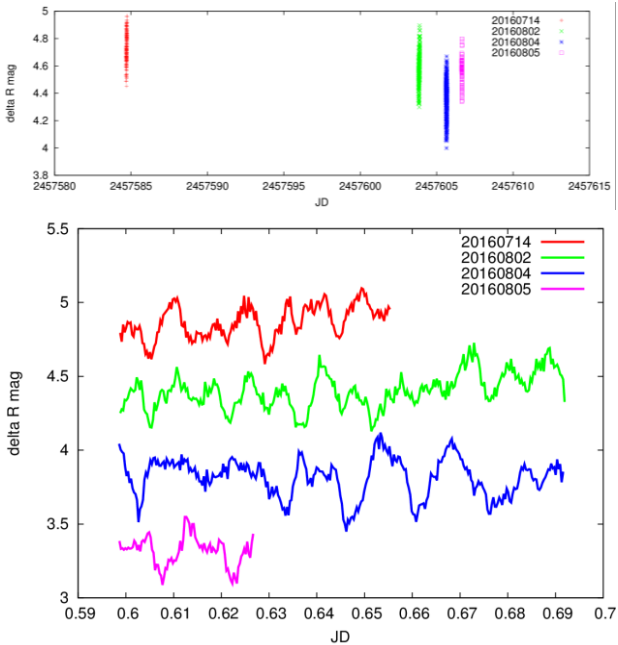

Figure 11. *(Top)* Light curve of FO Aqr including four nights in 2016. *(Bottom)* Same light curve but with proper time and magnitude offsets in order to show in detail the stellar variability.

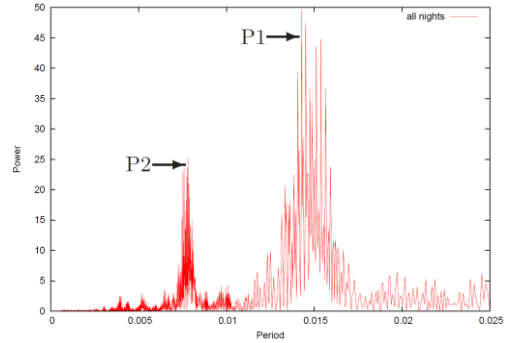

Figure 12. FO Aqr periodogram. The first period P1 = 0.01428987 days = 20.58 min and the second one P2 = 0.00781765 days = 11.25 min are shown.

### 3.4 Eclipsing binaries

Our cataclysmic variables monitoring program have observed approximately 9 hours (Fig. 1), distributed in three objects. In the following, we analyzed the data available for the eclipsing binary V357 Her observed in two nights (Table 1).

#### 3.4.1 V357 Her

V357 Her was discovered photographically about seventy years ago [30] and a period = 0.139725 days was derived. Considering the period and the shape of the light curve, the star was classified as an RR Lyrae-type variable, and reclassified later as a high-amplitude delta Scuti star [31]. Nevertheless, new CCD photometry lets to properly identify this star as a W UMa-Type eclipsing binary [32] with the real period being twice times the previous one claimed (P=0.28204111 days).

Figure 13 shows the FOV gathered with OAUNI around V357 Her on 2017/06/28 stacking the 2.8 hours sequence observed on this night. In this observation a R broadband filter was used. The light curve for V357 Her including the two observed nights (2017/06/08 and 2017/07/30) by OAUNI is shown in Figure 14-top. In order to show in detail the V357 variability, we added an offset (31.87 days) to the first night data trying to fit smoothly the resultant light curve with the second night data. This is shown in Figure 14-bottom where we can see approximately three-quarters of the nominal period for V357 Her. In this plot the minor minimum is observed on the first night data with the deepest minimum appearing on the second night.

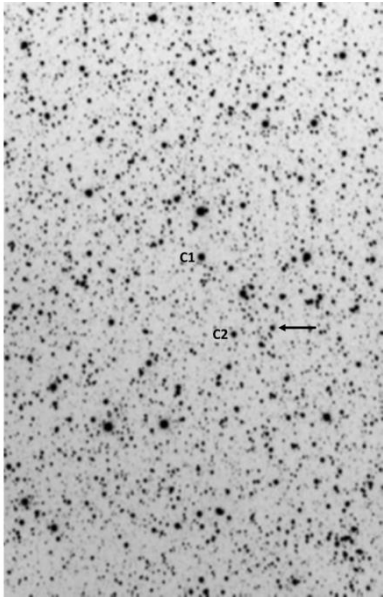

Figure 13. OAUNI stacked image (400 images×20s = 2.8h, including overhead) of V357 Her observed on 2017/06/28. North in top and East is left. The FOV shown is 12'×19'. The position of V357 Her is highlighted (arrow) along with the comparison stars (C1 and C2).

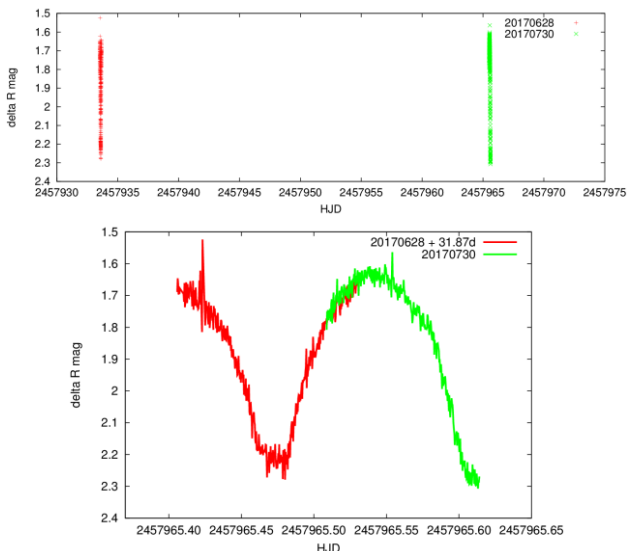

Figure 14. *(Top)* Light curve of V357 Her including two nights (2017/06/28-red and 2017/07/30-green). *(Bottom)* Same light curve but with 2017/06/30 data added 31.87d.

## 4   CONCLUSIONS

In this report we shown the actual status of the ongoing OAUNI program of short term period variable stars. More than ninety five hours of useful observations were carried out in this program during the 2016–2018 campaigns. The four subprograms which include monitoring of delta Scuti, rapidly oscillating stars, cataclysmic variables and eclipsing binaries have good quality data that will be publishing in everywhere. Examples of light curves were shown for each subprogram and it was verified the feasibility of use them to compute periodograms in order to find hidden periods of variability in each case.

Cases where the amplitude of variability if of order of tenths of magnitude were well resolved for objects so weak than ~13 mag and period of more than one hour. Our typical cadence between individual measurements of 20 sec (plus overhead) lets to plot well defined light curves. It was the case for the bright delta Scuti star CY Aqr (V = 10.4 mag) and the eclipsing binary V357 Her (V = 13.2). More challenged was the case of microvaribility of order the minutes, as in the cataclysmic variable FO Aqr observed in its in low state (V ~15.0–15.7). In this case, the hidden fundamental periods also were recovered by our light curves. Finally, cases with mmag variability were tested with success, as in the bright roAp star HD 217522 (V 7.6 mag), where the main 13.8 min period also was found. In general, we can conclude that our short term period variable stars program is in good accord with the photometric precision gathered with the OAUNI telescope and this give us confidence to continue this program in the next observational campaigns.

It is important to note that all the observational measurements in this program represent the first ones in their type carried out by professional peruvian astronomers at the central peruvian Andes. In this sense, the joint effort between the UNI and IGP to develop the observational peruvian astronomy is well addressed after the beginning of this project.

### ACKNOWLEDGEMENTS

The authors are grateful for the economic support from The World Academy of Sciences (TWAS), Rectorate and the Instituto General de Investigación (IGI) at UNI, and Concytec (Convenio 102-2015 Fondecyt). We are grateful to the Huancayo Observatory staff for the logistic support.

### REFERENCES

[1]  Pereyra A, Cori W, Meza E, Ricra J, Granda G 2012 *REVCIUNI* **15** 1 209
[2]  Pereyra A, Tello J, Meza E, Ricra J, Zevallos M 2015 *REVCIUNI* **18** 1 4
[3]  Pereyra A, Zevallos M, Ricra J, Tello J 2016 *Tecnia* **26** 2 20
[4]  Karttunen H, Kröger P, Oja H, Poutanen M, Donner K J 2007 *Fundamental Astronomy* (5th. Edition) Springer Ed. p 281
[5]  Catelan M, Smith H A 2015 *Pulsating stars* Wiley-VCH Verlag Ed. p 50
[6]  Breger M 2000 Delta Scuti and Related Stars - Astronomical Society of the Pacific Conference Series 210 3
[7]  Kurtz D W 1982 *MNRAS* **200** 807
[8]  Martinez P, Meintjes P, Ratcliff S J, Ebgelbrecht C 1998 Astronomy and Astrophysics 334 606
[9]  Connon Smith R 2007  arXiv:astro-ph/0701654
[10] Patterson J 1994 *PASP* **106** 209
[11] Downes R A, Webbink R F, Shara M M et al. 2006 *VizieR Online Data Catalog* **5123**
[12] Terrell D, Gross J, Cooney W R 2012 *AJ* **143** 99

[13] Hoffmeister C 1934 *Beob. Zirk. Astron. Nachr.* **16** 45
[14] Wiedemair C, Sterken C, Eenmäe T et al. 2016 *Journal of Astronomical Data* **22** 1
[15] Fang W J, Luo Z Q, Zhang X B et al. 2016 *Research in Astronomy and Astrophysics* **16** 96
[16] Cowall D E 2015 *Journal of the American Association of Variable Star Observers (JAAVSO)* **43** 201
[17] Sterken C, Wiedemair C, Tuvikene T et al. 2011 *Journal of Astronomical Data* **17**
[18] Hubscher J 2011 *Information Bulletin on Variable Stars* **5984** 1
[19] Hubscher J, Monninger G 2011 *Information Bulletin on Variable Stars* **5959** 1
[20] Scargle J D 1982 *ApJ* **263** 835
[21] Kurtz D W 1983 *MNRAS* **205** 3
[22] Kreidl T J, Kurtz D W, Bus S J et al. 1991 *MNRAS* **250** 477
[23] Medupe R, Kurtz D W, Elkin V G, Mguda Z, Mathys G 2015 *MNRAS* **446** 1347
[24] Osborne J, Mukai K 1989 *MNRAS* **238** 1233
[25] Littlefield C, Garnavich P, Kennedy M R et al. 2016 *ApJ* **833** 93
[26] Patterson J, Steiner J E 1983 *ApJ* **264** L61
[27] Littlefield C, Aadland E, Garnavich P, Kennedy M 2016 *The Astronomer's Telegram* **9225**
[28] Littlefield C, Garnavich P, Kennedy M R et al. 2016 *ApJ* **833** 93
[29] Littlefield C, Stiller R, Hambsch F J et al. 2018 *The Astronomer's Telegram* **11844**
[30] Goetz W, Wenzel W 1956 *Veroeffentlichungen der Sternwarte Sonneberg* **2** 279
[31] Rodríguez E, López-González M J, López de Coca P 2000 *A&As* **144** 469
[32] Branicki A, Pigulski A 2002 *Information Bulletin on Variable Stars* **5280** 1